\documentclass[journal=nalefd,manuscript=article]{achemso}

\usepackage[version=3]{mhchem} 
\usepackage[T1]{fontenc}       

\usepackage{amsmath}
\usepackage{amsfonts}
\usepackage{textcomp}
\usepackage[]{graphicx}
\usepackage{epstopdf}
\usepackage{mathrsfs}
\usepackage{amsbsy}
\usepackage{units}
\usepackage{MnSymbol}
\usepackage{lipsum}
\usepackage{color}
\usepackage{easybmat}

\usepackage{hyperref}
\usepackage{cleveref}

\crefformat{equation}{Eq.~(#2#1#3)}
\crefformat{figure}{Fig.~#2#1#3}

\Crefformat{equation}{Equation~(#2#1#3)}
\Crefformat{figure}{Figure~#2#1#3}

\author{Jingbo Sun}
\affiliation{Department of Electrical Engineering, The State University of New York at Buffalo, Buffalo, NY 14260, USA}
\altaffiliation{These authors have made equal contributions.}

\author{Salih Z. Silahli}
\affiliation{Department of Electrical Engineering, The State University of New York at Buffalo, Buffalo, NY 14260, USA}
\altaffiliation{These authors have made equal contributions.}

\author{Wiktor Walasik}
\affiliation{Department of Electrical Engineering, The State University of New York at Buffalo, Buffalo, NY 14260, USA}
\altaffiliation{These authors have made equal contributions.}

\author{Qi Li}

\affiliation{Department of Chemical and Biological Engineering, The State University of New York, University at Buffalo, Buffalo, NY 14260, USA}

\author{Eric Johnson}
\affiliation{Department of Electrical Engineering and Computer Science, Clemson University, Clemson, SC 29634, USA}

\author{Natalia M. Litchinitser}
\affiliation{Department of Electrical Engineering, The State University of New York at Buffalo, Buffalo, NY 14260, USA}
\email{natashal@buffalo.edu}

\title{Orbital angular momentum beam instabilities in engineered nonlinear colloidal media}

\abbreviations{IR,NMR,UV}

\begin{document}

\begin{tocentry}
	\centering
	\includegraphics[height=3.5cm,clip=true,trim= 0 0 0 0]{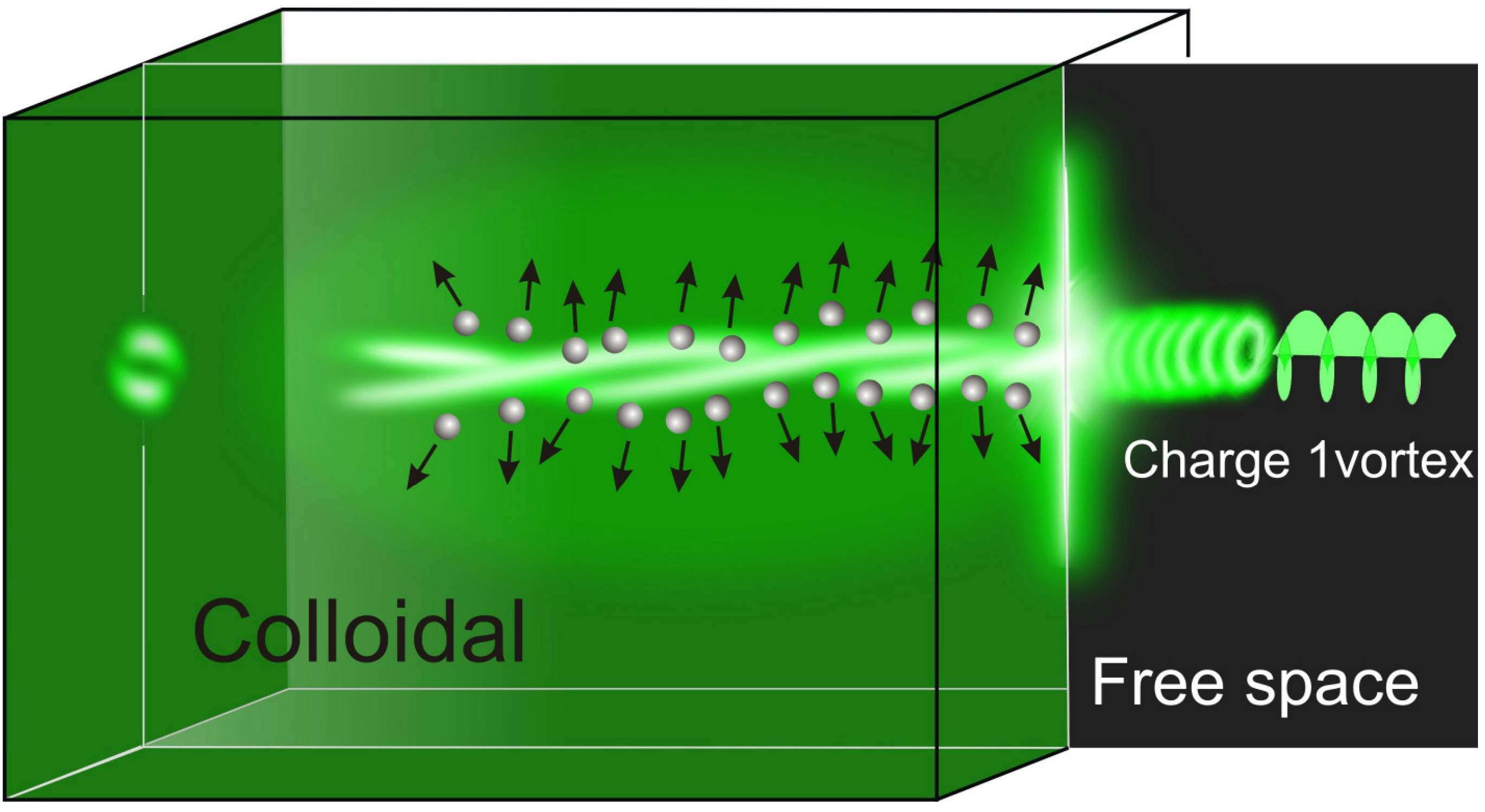}

\end{tocentry}

\begin{abstract}
  In this letter, we experimentally demonstrate the evolution of the optical vortex beams of different topological charges propagating in engineered nano-colloidal suspension of negative polarizability with saturable nonlinearities. Due to the high power of the incident beam, the modulation instability leads to an exponential growth of weak perturbations and thus splits the original vortex beam into a necklace beam consisting of several bright spots. The number of observed bright spots is intrinsically determined by the topological charge of the incident beam and agrees well with the predictions of our linear stability analysis and numerical simulations. Besides contributing to the fundamental science of light-matter interactions in engineered soft-matter media, this work opens new opportunities for dynamic optical manipulation and transmission of light through scattering media as well as formation of complex optical patterns and light filamentation in naturally existing colloids such as fog and clouds.
\end{abstract}


Almost a century ago, the work of Einstein~\cite{einstein1956investigations} and Perrin~\cite{perrin1916} laid the foundations of modern physics of colloids---liquids containing structures on the scale of roughly 10~nm to 1~$\mu$m that are stable against sedimentation. Since then colloids with well-defined particle size, shape and interaction lengths have been widely used as model systems in fundamental studies of statistical physics phenomena~\cite{hunter2001foundations}, phase transitions and optical trapping~\cite{Dholakia2011} to name a few. Propagation of light beams through some common colloidal media such as fog, clouds, smoke, paints, and milk finds increasingly important applications in science and technology, ranging from optical bar-coding for applications in genomics, proteomics and drug discovery~\cite{B200038P}, free-space communication technologies~\cite{Wu:08} and weather control~\cite{Rohwetter2010} to security and defense~\cite{mcaulay2011military}.

Recent progress in the development of artificial materials, or metamaterials, with fundamentally new physical properties opens new opportunities for tailoring the properties of colloids. Metamaterials are built of resonant elements with dimensions much smaller than the wavelength of light, sometimes referred to as meta-atoms, enabling light-matter interactions that are difficult or impossible to realize using naturally available materials. A majority of photonic metamaterials that have been demonstrated to date were solid-state materials. However, the concept of meta-atoms can be extended further to realize artificial media with novel electromagnetic properties in liquid~\cite{ADMA:ADMA201670049} or gaseous~\cite{Kudyshev2013a} phases at frequencies ranging from microwave to visible. In particular, at optical frequencies, engineered colloidal suspensions offer as a promising platform for engineering polarizabilities and realization of large and tunable nonlinearities. Recent studies have shown that the nonlinearity of colloidal suspensions has exponential character and can be either supercritical, in case of particles with positive polarizability, or saturable, for negative polarizability particles~\cite{El-Ganainy:07,El-Ganainy:07b,PhysRevLett.111.218302,Kelly:16}. 

To date, such engineered colloidal systems have been studied using simple Gaussian beams. However, recent progress in structuring amplitude and phase properties of optical beams opens new remarkable opportunities for manipulating and controlling light-matter interactions in such engineered media. Compared to the conventionally used Gaussian beams, optical vortices that, are characterized by the doughnut-shaped intensity profile and a helical phase front, offer even more degrees of freedom for optical trapping~\cite{95302849a5214942adf2c240ebe9bee6} or imaging applications~\cite{Xie:13}. Optical vortices can be used to trap and circulate colloidal particles, constituting a model test-bed for studying many-body hydrodynamic coupling and instabilities in mesoscopic, many-particle systems with potential applications in lab-on-a-chip systems~\cite{Dholakia2011,Ladavac:04,Lee:06,Reichert2006Hydro-9458}.

In this letter, we experimentally investigate the evolution of the optical vortex beams of different topological charges in engineered nano-colloidal suspensions with saturable nonlinearities, in which the particles with negative polarizability are repelled away from the high-intensity region. As the high-intensity vortex beam propagates in such a medium, the modulation instability (MI) phenomenon leads to an exponential growth of weak perturbations. As we predicted in our linear stability analysis and numerical simulations~\cite{Silahli:15}, the perturbations with an orbital angular momentum (OAM) of a particular charge is amplified leading to the formation of a necklace beam with a well-defined number of peaks. The experimental results are in excellent agreement with the analytical and numerical predictions. Besides contributing to the fundamental science of light-matter interactions in engineered soft-matter media, our work might bring about new possibilities for dynamic optical manipulation and transmission of light through scattering media as well as formation of complex optical patterns and light filamentation~\cite{PhysRevLett.95.193901,Vincotte2006163,Walasik2017} in naturally existing colloids such as fog and clouds.


\begin{figure}[!t]
	\centering
	\includegraphics[width=0.7\textwidth,clip=true,trim= 0 0 0 0]{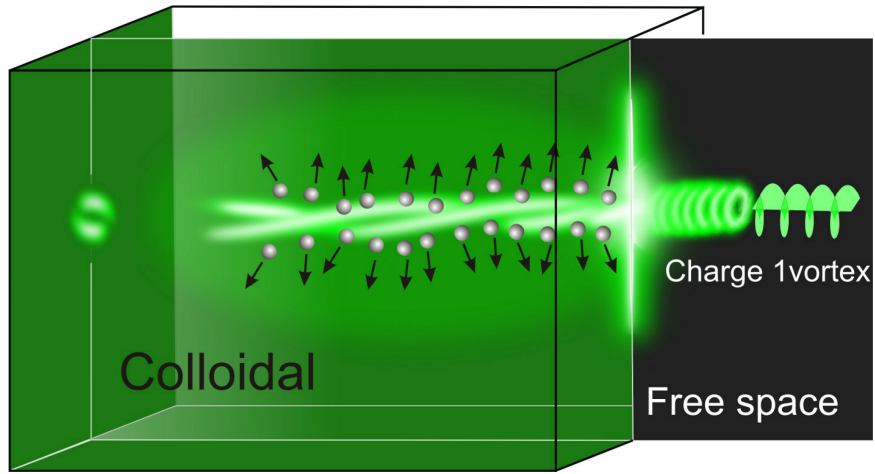}
	\caption{Propagation of a charge one optical vortex beam in a colloidal solution with negative polarizability. In free space, the helical wave front (right) and the doughnut intensity profile of the beam (left) are schematically shown. Inside the colloidal medium, the input vortex beam transforms into a rotating necklace beam. The repulsion of the particles in the path of the high intensity beam leads to a local nonlinear index change.}
	\label{fig:scheme}
\end{figure}

Let us consider an optical vortex beam propagating along the $z$-direction in a nano-colloidal system consisting of dielectric particles with refractive index $n_p$ lower than the refractive index of the background medium $n_b$. If $n_p<n_b$, the colloidal suspension has a negative polarizability, as schematically illustrated in \cref{fig:scheme}, the nano-particles are driven away from the high intensity region of the beam, resulting in a change of the local refractive index in the suspension, which exhibits a focusing nonlinearity. For large input intensities, the beam becomes unstable due to the well-known phenomenon of MI. This effect reveals itself as the exponential growth of weak perturbations or noise in the presence of an intense pump beam propagating in a nonlinear medium. As a result of the MI, the original vortex beam of a doughnut shape may split into a necklace-like beam with several bright spots, whose number is intrinsically determined by the topological charge of the vortex beam. This process is described by the nonlinear Schr\"odinger equation (NLSE) (see Methods). Following the standard linear stability analysis~\cite{Silahli:15,Vincotte2006163} we assume that the high-intensity optical beam with a topological vortex charge $m\in\mathbb{Z}$ is accompanied by an azimuthal perturbation:
\begin{equation}
E(\theta,z) = \left[ |E_0| + a_1 e^{-i(M\theta + \mu z)} + a_2^* e^{i(M\theta + \mu^* z)} \right] e^{i(m\theta + \lambda z)}
\end{equation}
where $E_0 = E(r=r_m,z=0)$ is the electric field amplitude of the rotationally invariant steady state solution of \cref{eq:NLSE} (see Methods) with charge $m$, taken at the average radius $r_m$, $a_1$, $a_2$ are the amplitudes of the small perturbations, and $M\in \mathbb{Z}$ is the deviation from $m$ of the perturbation charge. The topological charge of the perturbation is given by $m \pm M$,  $\lambda$ is the propagation constant of the steady state solution, and $\mu$ is the propagation constant correction for the perturbation.
The linear stability analysis allows us to calculate the MI gain for the perturbations with the charge $m \pm M$ imposed atop the main beam with the charge $m$ and the corresponding averaged radius $r_m$. The gain is given by~\cite{Silahli:15}:

\begin{equation}
\mathrm{Im}(\mu) = g_m(M) = \frac{M}{2 k_0 n_b r_m} \times \mathrm{Im}\sqrt{ \frac{M^2}{r_m^2} - \frac{|\alpha|}{2 k_B T L^2} |E_0|^2 \exp\left(\frac{\alpha}{2 k_B T} |E_0|^2\right) }
\label{eq:gain}
\end{equation}
where $L^2 = (2 k_0^2 n_b |n_p - n_b| V_p \rho_0)^{-1} $. Here, the particle polarizability is denoted by $\alpha$, and $k_B T$ is the thermal energy, with the Boltzmann constant $k_B$ and at temperature $T$. $V_p$ is the volume of a particle, $\rho_0$ is the unperturbed particle concentration, $k_0=\frac{2\pi}{\lambda_0}$ is the wave number, and $\lambda_0$ is the free-space wavelength.

\begin{figure}[!b]
	\centering
	\includegraphics[width=\textwidth,clip=true,trim= 90 0 90 0]{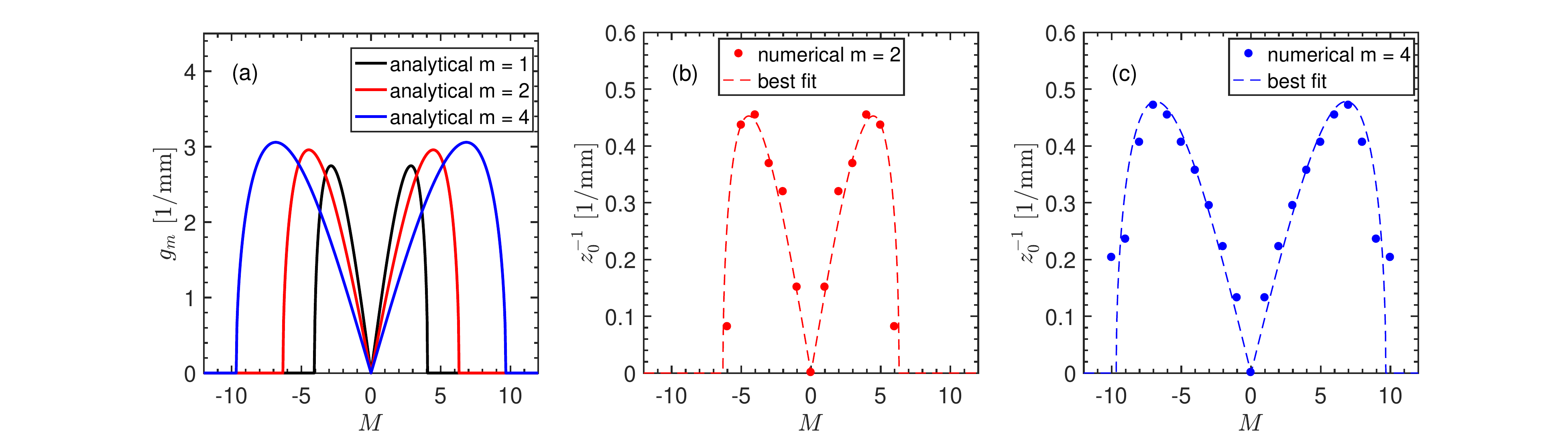}
	\caption{Azimuthal modulation instability gain. (a) Analytically computed instability gain $g_m(M)$ as a function of the perturbation azimuthal index deviation $M$, for negative polarizability particle-based systems for different topological charges $m$ of the initial steady-state vortex solution. (b)--(c) Inverse of the beam breakup distance recorded in numerical simulations of seeded MI. Dashed lines show analytical curves with rescaled magnitude that help guide the eye.}
	\label{fig:analytical}
\end{figure}

\Cref{eq:gain} is used in the following to predict the MI gain for vortices propagating in the nano-colloidal media. We study the propagation of light with the free-space wavelength $\lambda_0 = 532$~nm, the negative-polarizability suspension made of low refractive index polytetrafluoroethylene (PTFE) particles ($n_p=1.35$) dispersed in glycerin water ($n_b = 1.44$) with the volume filling fraction of $\rho_0 = 0.7\%$. The radius of the particles is assumed to be $150$~nm and the experiments were performed at room temperature. For this set of the parameters, \cref{fig:analytical}(a) shows the gain curves $g_m(M)$ as a function of the perturbation azimuthal index deviation $M$ for different values of the vortex charge $m$. The analytical predictions are only valid when the perturbation intensity is significantly lower than that of the main beam. Above this limit the dynamics of MI has to be studied using numerical simulations of a three-dimensional NLSE (see Methods). 

In order to confirm the analytical prediction for the number of maxima and the shape of the gain curves, we have numerically solved the NLSE in the absence of scattering losses ($\sigma=0$) using the split-step Fourier method~\cite{Feit:78,doi:10.1063/1.328442}. First, based on the theoretical predictions, we have found the parameters of the stationary vortex solitons with charges $m=2$ and $m=4$. We have numerically confirmed that for a given parameters of the stable beam (power and average radius $r_m$) the vortex propagates in a stable manner, provided that the medium is lossless. Addition of the random noise on top of the stable solution resulted in the MI induced beam breakup into a necklace beam with the number of maxima predicted by the analytical results ($N=4$ for the main vortex charge $m=2$, and $N=7$ for the main vortex charge $m=4$). 
Simulations of the MI allow us to determine the rate at which the pattern with $N$ maxima grows. In order to seed the growth of a pattern with $N$ maxima, we add only the perturbation of charges $m \pm M = N$. The distance $z_0$ at which the pattern with $N$ maxima emerges is inversely proportional to the  modulation gain $g_m(M)$. The distance $z_0$ is read from the light intensity maps $I(r, \theta, z)$, and its choice is somewhat arbitrary. We have chosen $z_0$ to be the distance at which the contrast between the $N$ maxima and the minima in between them is the highest. The results of $1/z_0$ for main vortex charges $m=2$ and $m=4$ are shown in \cref{fig:analytical}(b), (c). We can see a great agreement with the rescaled analytical curves showing the MI gain $1/z_0$.

\begin{figure}[!t]
	\centering
	\includegraphics[width=0.8\textwidth,clip=true,trim= 0 0 0 0]{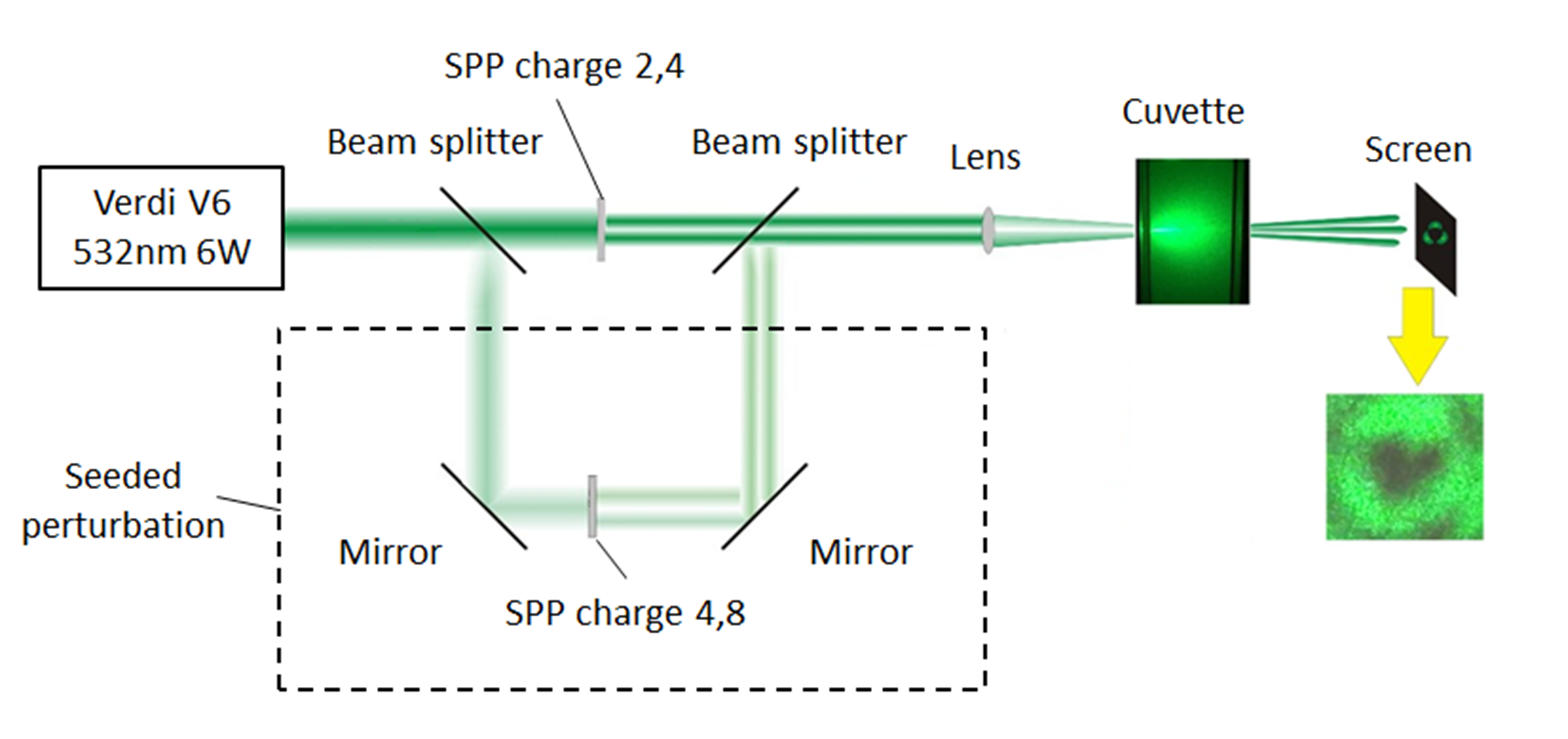}
	\caption{Experimental setup used to study (seeded) modulation instability of vortex beams in colloidal media. Collimated beam from Verdi V6 laser ($\lambda_0 =532$~nm) is initially split into two beams using beam splitters with reflectivity varying in the range from $0.6\%$ to $8\%$ of the total power. The high intensity beam is transmitted through a spiral phase plate (SPP) to generate the main vortex beam with lower charge. In the seeded configuration, the low intensity beam is transmitted through a SPP with a higher charge to generate the perturbation beam. The beams are then recombined at the second beam splitter and focused onto the cuvette by a lens. The longitudinal beam profile inside the cuvette and the transverse beam profile behind the cuvette are recorded by a camera and shown in the insets.}
	\label{fig:setup}
\end{figure}

\begin{figure}[!t]
	\centering
	\includegraphics[width=0.7\textwidth,clip=true,trim= 0 0 0 0]{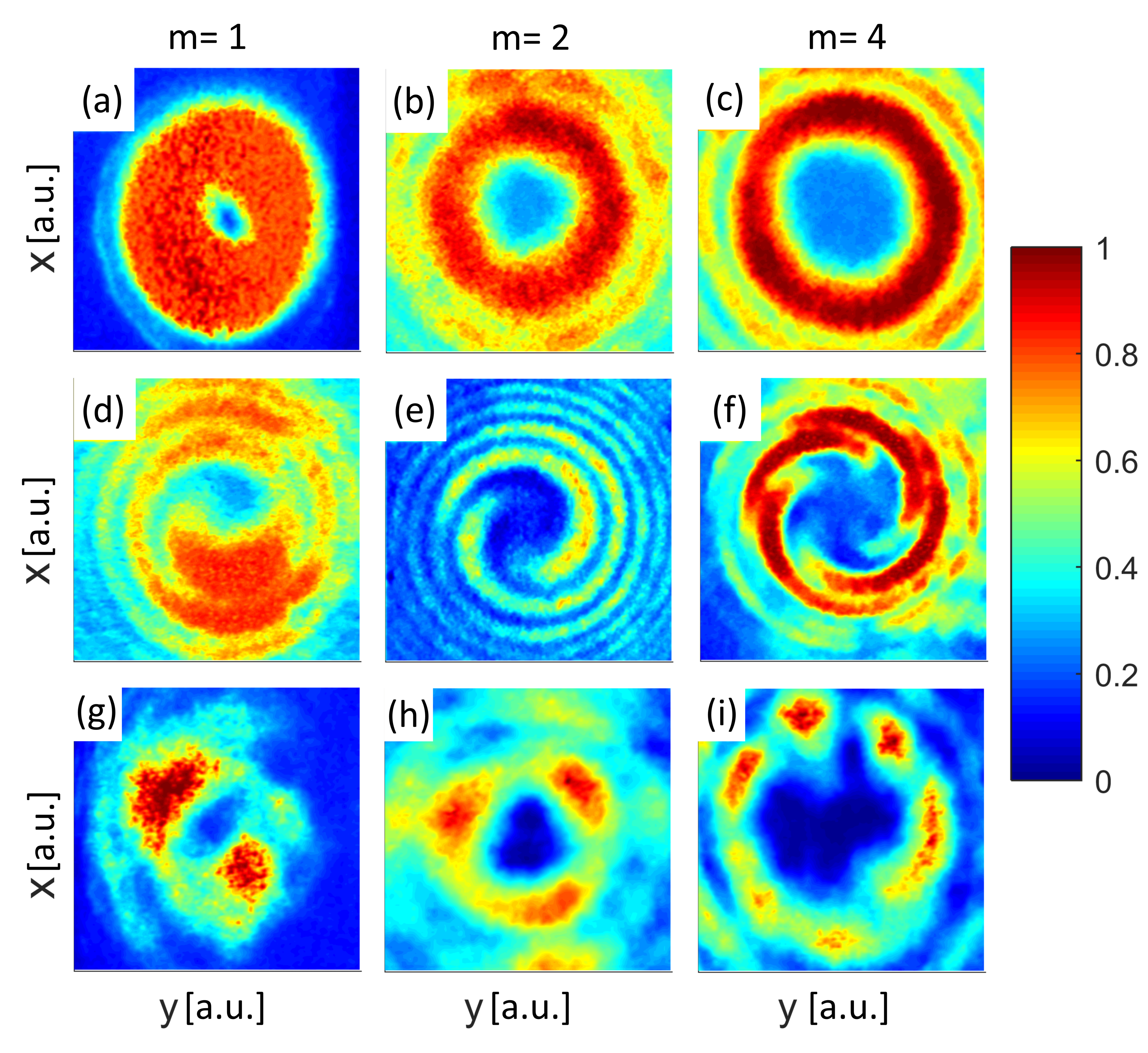}
	\caption{Experimental results showing the formation of the necklace beam from an initial vortex beam propagating in a nonlinear colloidal suspension with negative polarizability particles. (a)--(c) Intensity profiles of the incident vortex beams of charges 1, 2, and 4. (d)--(f) Interference patterns corresponding to vortex beams with topological charges in (a)--(c), respectively. (g)--(i) Intensity distributions of the resulting necklace beams after the propagation in the colloidal medium corresponding to the incident beams (a)--(c), respectively.}
	\label{fig:res}
\end{figure}

In our experiments, the beam from a 532~nm, 6~W, continuous wave Coherent Verdi 6 laser was first converted into an optical vortex beam using a spiral phase plate and then focused inside a 10-mm-long cuvette filled with the colloidal suspension consisting of PTFE particles [Laurel, Ultraflon AD-10] dispersed in glycerin/water solution (3:1, v/v), as shown in \cref{fig:setup}. The filling ratio of the PTFE is 0.7\%. Since the refractive index of the PTFE particles is lower than that of glycerin water~\cite{Silahli:15}, the particles have negative polarizability. First, we observed MI growing from noise, i.e. without a well-defined perturbation. \Cref{fig:res}(a)--(c) shows different optical vortices of charge 1, 2, and 4, generated using the spiral phase plates.

Interference experiments were performed to confirm the topological charges of the generated vortex beams, as shown in \cref{fig:res}(d)--(f). Due to the MI, the original doughnut-shaped beam after passing through the colloidal suspension splits into several bright spots, depending on its initial charge. Here, we performed two series of experiments with and without the seed, as shown in \cref{fig:setup}. For the incident beam with $m=1$ which was directly focused into the cuvette without adding any induced perturbation, the beam after passing through the colloidal solution splits into 2 bright spots, as shown in \cref{fig:res}(g). In the case of seeded (or induced) MI, we investigated the propagation of the vortex beams of $m=2$ and $m=4$. Firstly, we explored the case of focused beam with $m=2$ in the presence of weak seeded perturbations. Perturbations of charge 4 and charge 8 were added separately and then together to the main beam. The intensity ratio between the perturbations and the main was adjusted from 0 to 3\%. By carefully testing all these cases with different perturbation charges and intensities, we find that in such a competition between the perturbations originating from the noise and those seeded by the low intensity beam, the final beam pattern on the screen always shows 3 maxima. This result is qualitatively consistent with our analytical predictions, revealing the fact that only the perturbation with the charge close to the maximum of the gain curve is amplified, as it grows faster than other perturbations, even if they are seeded. Secondly, a similar test was performed for the vortex with charge 4 in the presence of the perturbation with charge 8. The final pattern observed on the screen shows a necklace beam including 7 maxima, which also corresponds to the maximum of the gain curve shown in \cref{fig:analytical}(c).


In summary, we have experimentally and numerically studied both seeded and unseeded modulation instability in colloidal suspensions of negative polarizability nano-particles. The experimental results are in good agreement with the numerical predictions. In particular, in the case of seeded modulation instability, the observed necklace beam patterns were identical with the patterns obtained without the seed. This shows that the perturbation with the largest growth rate predicted in the analytical and numerical calculations prevails over all the other perturbations introduced to the beam either through the noise or as a seeded perturbation. These results are likely to enable a new platform for fundamental studies of nonlinear optical phenomena in engineered media as well as for imaging and light manipulation in scattering media, such as biological and chemical systems.

\section{Methods}

The nonlinear Schr\"odinger equation governing the evolution of the slowly varying electric field envelope $E$ can be written as~\cite{El-Ganainy:07,Silahli:15}:
\begin{equation}
i \frac{\partial E}{\partial z} + \frac{1}{2 k_0 n_b} \nabla^2_{\perp} E + k_0 (n_b - n_p) V_p \rho_0 e^{\frac{\alpha}{4 k_B T}|E|^2} E + \frac{i}{2} \sigma \rho_0 e^{\frac{\alpha}{4 k_B T}|E|^2} E = 0
\label{eq:NLSE}
\end{equation}
where $\nabla^2_{\perp} = \frac{1}{r}\frac{\partial }{\partial r}\left( r \frac{\partial }{\partial r} \right) +\frac{1}{r^2}\frac{\partial^2 }{\partial \theta^2}$ is the transverse Laplacian. The particle polarizability is denoted by $\alpha$, and $k_B T$ is the thermal energy, with the Boltzmann constant $k_B$ and at temperature $T$, $V_p$ is the volume of a particle, $\rho_0$ is the unperturbed particle concentration, $\sigma$ is the scattering cross-section, $k_0=\frac{2\pi}{\lambda_0}$ is the wave number, and $\lambda_0$ is the free-space wavelength. This equation was analyzed in detail using the linear stability analysis~\cite{Silahli:15} and solved numerically using the split-step Fourier algorithm.

\begin{acknowledgement}
	
	Army Research Office [W911NF-11-1-0297, W911NF-15-1-0146]. 
	
	The authors thank professor D. Christodoulides from University of Central Florida fruitful discussions.

\end{acknowledgement}

%
%
%


\begin{mcitethebibliography}{26}
	\providecommand*\natexlab[1]{#1}
	\providecommand*\mciteSetBstSublistMode[1]{}
	\providecommand*\mciteSetBstMaxWidthForm[2]{}
	\providecommand*\mciteBstWouldAddEndPuncttrue
	{\def\EndOfBibitem{\unskip.}}
	\providecommand*\mciteBstWouldAddEndPunctfalse
	{\let\EndOfBibitem\relax}
	\providecommand*\mciteSetBstMidEndSepPunct[3]{}
	\providecommand*\mciteSetBstSublistLabelBeginEnd[3]{}
	\providecommand*\EndOfBibitem{}
	\mciteSetBstSublistMode{f}
	\mciteSetBstMaxWidthForm{subitem}{(\alph{mcitesubitemcount})}
	\mciteSetBstSublistLabelBeginEnd
	{\mcitemaxwidthsubitemform\space}
	{\relax}
	{\relax}
	
	\bibitem[Einstein(1926)]{einstein1956investigations}
	Einstein,~A. \emph{Investigations on the Theory of the Brownian Movement};
	Dover, New York, 1926\relax
	\mciteBstWouldAddEndPuncttrue
	\mciteSetBstMidEndSepPunct{\mcitedefaultmidpunct}
	{\mcitedefaultendpunct}{\mcitedefaultseppunct}\relax
	\EndOfBibitem
	\bibitem[Perrin(1916)]{perrin1916}
	Perrin,~J. \emph{Les Atomes}; Constable, London, 1916\relax
	\mciteBstWouldAddEndPuncttrue
	\mciteSetBstMidEndSepPunct{\mcitedefaultmidpunct}
	{\mcitedefaultendpunct}{\mcitedefaultseppunct}\relax
	\EndOfBibitem
	\bibitem[Hunter(2001)]{hunter2001foundations}
	Hunter,~R.~J. \emph{Foundations of colloid science}; Oxford University Press,
	2001\relax
	\mciteBstWouldAddEndPuncttrue
	\mciteSetBstMidEndSepPunct{\mcitedefaultmidpunct}
	{\mcitedefaultendpunct}{\mcitedefaultseppunct}\relax
	\EndOfBibitem
	\bibitem[Dholakia and {\v{C}}i{\v{z}}m{\'a}r(2011)Dholakia, and
	{\v{C}}i{\v{z}}m{\'a}r]{Dholakia2011}
	Dholakia,~K.; {\v{C}}i{\v{z}}m{\'a}r,~T. \emph{Nat. Photon.} \textbf{2011},
	\emph{5}, 335--342\relax
	\mciteBstWouldAddEndPuncttrue
	\mciteSetBstMidEndSepPunct{\mcitedefaultmidpunct}
	{\mcitedefaultendpunct}{\mcitedefaultseppunct}\relax
	\EndOfBibitem
	\bibitem[Battersby \latin{et~al.}(2002)Battersby, Lawrie, Johnston, and
	Trau]{B200038P}
	Battersby,~B.~J.; Lawrie,~G.~A.; Johnston,~A. P.~R.; Trau,~M. \emph{Chem.
		Commun.} \textbf{2002}, 1435--1441\relax
	\mciteBstWouldAddEndPuncttrue
	\mciteSetBstMidEndSepPunct{\mcitedefaultmidpunct}
	{\mcitedefaultendpunct}{\mcitedefaultseppunct}\relax
	\EndOfBibitem
	\bibitem[Wu \latin{et~al.}(2008)Wu, Hajjarian, and Kavehrad]{Wu:08}
	Wu,~B.; Hajjarian,~Z.; Kavehrad,~M. \emph{Appl. Opt.} \textbf{2008}, \emph{47},
	3168--3176\relax
	\mciteBstWouldAddEndPuncttrue
	\mciteSetBstMidEndSepPunct{\mcitedefaultmidpunct}
	{\mcitedefaultendpunct}{\mcitedefaultseppunct}\relax
	\EndOfBibitem
	\bibitem[Rohwetter \latin{et~al.}(2010)Rohwetter, Kasparian, Stelmaszczyk, Hao,
	Henin, Lascoux, Nakaema, Petit, Quei{\ss}er, Salame, Salmon, W\"oste, and
	Wolf]{Rohwetter2010}
	Rohwetter,~P.; Kasparian,~J.; Stelmaszczyk,~K.; Hao,~Z.; Henin,~S.;
	Lascoux,~N.; Nakaema,~W.~M.; Petit,~Y.; Quei{\ss}er,~M.; Salame,~R.;
	Salmon,~E.; W\"oste,~L.; Wolf,~J.-P. \emph{Nat. Photon.} \textbf{2010},
	\emph{4}, 451--456\relax
	\mciteBstWouldAddEndPuncttrue
	\mciteSetBstMidEndSepPunct{\mcitedefaultmidpunct}
	{\mcitedefaultendpunct}{\mcitedefaultseppunct}\relax
	\EndOfBibitem
	\bibitem[McAulay(2011)]{mcaulay2011military}
	McAulay,~A.~D. \emph{Military laser technology for defense: Technology for
		revolutionizing 21st century warfare}; John Wiley \& Sons, New Jersey,
	2011\relax
	\mciteBstWouldAddEndPuncttrue
	\mciteSetBstMidEndSepPunct{\mcitedefaultmidpunct}
	{\mcitedefaultendpunct}{\mcitedefaultseppunct}\relax
	\EndOfBibitem
	\bibitem[Liu \latin{et~al.}(2016)Liu, Fan, Padilla, Powell, Zhang, and
	Shadrivov]{ADMA:ADMA201670049}
	Liu,~M.; Fan,~K.; Padilla,~W.; Powell,~D.~A.; Zhang,~X.; Shadrivov,~I.~V.
	\emph{Adv. Mater.} \textbf{2016}, \emph{28}, 1525--1525\relax
	\mciteBstWouldAddEndPuncttrue
	\mciteSetBstMidEndSepPunct{\mcitedefaultmidpunct}
	{\mcitedefaultendpunct}{\mcitedefaultseppunct}\relax
	\EndOfBibitem
	\bibitem[Kudyshev \latin{et~al.}(2013)Kudyshev, Richardson, and
	Litchinitser]{Kudyshev2013a}
	Kudyshev,~Z.~A.; Richardson,~M.~C.; Litchinitser,~N.~M. \emph{Nat. Commun.}
	\textbf{2013}, \emph{4}, 2557\relax
	\mciteBstWouldAddEndPuncttrue
	\mciteSetBstMidEndSepPunct{\mcitedefaultmidpunct}
	{\mcitedefaultendpunct}{\mcitedefaultseppunct}\relax
	\EndOfBibitem
	\bibitem[El-Ganainy \latin{et~al.}(2007)El-Ganainy, Christodoulides,
	Musslimani, Rotschild, and Segev]{El-Ganainy:07}
	El-Ganainy,~R.; Christodoulides,~D.~N.; Musslimani,~Z.~H.; Rotschild,~C.;
	Segev,~M. \emph{Opt. Lett.} \textbf{2007}, \emph{32}, 3185--3187\relax
	\mciteBstWouldAddEndPuncttrue
	\mciteSetBstMidEndSepPunct{\mcitedefaultmidpunct}
	{\mcitedefaultendpunct}{\mcitedefaultseppunct}\relax
	\EndOfBibitem
	\bibitem[El-Ganainy \latin{et~al.}(2007)El-Ganainy, Christodoulides, Rotschild,
	and Segev]{El-Ganainy:07b}
	El-Ganainy,~R.; Christodoulides,~D.~N.; Rotschild,~C.; Segev,~M. \emph{Opt.
		Express} \textbf{2007}, \emph{15}, 10207--10218\relax
	\mciteBstWouldAddEndPuncttrue
	\mciteSetBstMidEndSepPunct{\mcitedefaultmidpunct}
	{\mcitedefaultendpunct}{\mcitedefaultseppunct}\relax
	\EndOfBibitem
	\bibitem[Man \latin{et~al.}(2013)Man, Fardad, Zhang, Prakash, Lau, Zhang,
	Heinrich, Christodoulides, and Chen]{PhysRevLett.111.218302}
	Man,~W.; Fardad,~S.; Zhang,~Z.; Prakash,~J.; Lau,~M.; Zhang,~P.; Heinrich,~M.;
	Christodoulides,~D.~N.; Chen,~Z. \emph{Phys. Rev. Lett.} \textbf{2013},
	\emph{111}, 218302\relax
	\mciteBstWouldAddEndPuncttrue
	\mciteSetBstMidEndSepPunct{\mcitedefaultmidpunct}
	{\mcitedefaultendpunct}{\mcitedefaultseppunct}\relax
	\EndOfBibitem
	\bibitem[Kelly \latin{et~al.}(2016)Kelly, Ren, Samadi, Bezryadina,
	Christodoulides, and Chen]{Kelly:16}
	Kelly,~T.~S.; Ren,~Y.-X.; Samadi,~A.; Bezryadina,~A.; Christodoulides,~D.;
	Chen,~Z. \emph{Opt. Lett.} \textbf{2016}, \emph{41}, 3817--3820\relax
	\mciteBstWouldAddEndPuncttrue
	\mciteSetBstMidEndSepPunct{\mcitedefaultmidpunct}
	{\mcitedefaultendpunct}{\mcitedefaultseppunct}\relax
	\EndOfBibitem
	\bibitem[Rubinsztein-Dunlop \latin{et~al.}(2017)Rubinsztein-Dunlop, Forbes,
	Berry, Dennis, Andrews, Mansuripur, Denz, Alpmann, Banzer, Bauer, Karimi,
	Marrucci, Padgett, Ritsch-Marte, Litchinitser, Bigelow, Rosales-Guzmán,
	Belmonte, Torres, Neely, Baker, Gordon, Stilgoe, Romero, White, Fickler,
	Willner, Xie, McMorran, and Weiner]{95302849a5214942adf2c240ebe9bee6}
	Rubinsztein-Dunlop,~H. \latin{et~al.}  \emph{J. Opt.} \textbf{2017}, \emph{19},
	013011\relax
	\mciteBstWouldAddEndPuncttrue
	\mciteSetBstMidEndSepPunct{\mcitedefaultmidpunct}
	{\mcitedefaultendpunct}{\mcitedefaultseppunct}\relax
	\EndOfBibitem
	\bibitem[Xie \latin{et~al.}(2013)Xie, Liu, Jin, Santangelo, and Xi]{Xie:13}
	Xie,~H.; Liu,~Y.; Jin,~D.; Santangelo,~P.~J.; Xi,~P. \emph{J. Opt. Soc. Am. A}
	\textbf{2013}, \emph{30}, 1640--1645\relax
	\mciteBstWouldAddEndPuncttrue
	\mciteSetBstMidEndSepPunct{\mcitedefaultmidpunct}
	{\mcitedefaultendpunct}{\mcitedefaultseppunct}\relax
	\EndOfBibitem
	\bibitem[Ladavac and Grier(2004)Ladavac, and Grier]{Ladavac:04}
	Ladavac,~K.; Grier,~D.~G. \emph{Opt. Express} \textbf{2004}, \emph{12},
	1144--1149\relax
	\mciteBstWouldAddEndPuncttrue
	\mciteSetBstMidEndSepPunct{\mcitedefaultmidpunct}
	{\mcitedefaultendpunct}{\mcitedefaultseppunct}\relax
	\EndOfBibitem
	\bibitem[Lee \latin{et~al.}(2006)Lee, Garc\'{e}s-Ch\"{a}vez, and
	Dholakia]{Lee:06}
	Lee,~W.~M.; Garc\'{e}s-Ch\"{a}vez,~V.; Dholakia,~K. \emph{Opt. Express}
	\textbf{2006}, \emph{14}, 7436--7446\relax
	\mciteBstWouldAddEndPuncttrue
	\mciteSetBstMidEndSepPunct{\mcitedefaultmidpunct}
	{\mcitedefaultendpunct}{\mcitedefaultseppunct}\relax
	\EndOfBibitem
	\bibitem[Reichert(2006)]{Reichert2006Hydro-9458}
	Reichert,~M. Hydrodynamic Interactions in Colloidal and Biological Systems.
	Ph.D.\ thesis, Universit\"at Konstanz, Konstanz, 2006\relax
	\mciteBstWouldAddEndPuncttrue
	\mciteSetBstMidEndSepPunct{\mcitedefaultmidpunct}
	{\mcitedefaultendpunct}{\mcitedefaultseppunct}\relax
	\EndOfBibitem
	\bibitem[Silahli \latin{et~al.}(2015)Silahli, Walasik, and
	Litchinitser]{Silahli:15}
	Silahli,~S.~Z.; Walasik,~W.; Litchinitser,~N.~M. \emph{Opt. Lett.}
	\textbf{2015}, \emph{40}, 5714--5717\relax
	\mciteBstWouldAddEndPuncttrue
	\mciteSetBstMidEndSepPunct{\mcitedefaultmidpunct}
	{\mcitedefaultendpunct}{\mcitedefaultseppunct}\relax
	\EndOfBibitem
	\bibitem[Vin\ifmmode~\mbox{\c{c}}\else \c{c}\fi{}otte and
	Berg\'e(2005)Vin\ifmmode~\mbox{\c{c}}\else \c{c}\fi{}otte, and
	Berg\'e]{PhysRevLett.95.193901}
	Vin\ifmmode~\mbox{\c{c}}\else \c{c}\fi{}otte,~A.; Berg\'e,~L. \emph{Phys. Rev.
		Lett.} \textbf{2005}, \emph{95}, 193901\relax
	\mciteBstWouldAddEndPuncttrue
	\mciteSetBstMidEndSepPunct{\mcitedefaultmidpunct}
	{\mcitedefaultendpunct}{\mcitedefaultseppunct}\relax
	\EndOfBibitem
	\bibitem[Vin\ifmmode~\mbox{\c{c}}\else \c{c}\fi{}otte and
	Berg\'e(2006)Vin\ifmmode~\mbox{\c{c}}\else \c{c}\fi{}otte, and
	Berg\'e]{Vincotte2006163}
	Vin\ifmmode~\mbox{\c{c}}\else \c{c}\fi{}otte,~A.; Berg\'e,~L. \emph{Physica D:
		Nonlinear Phenomena} \textbf{2006}, \emph{223}, 163 -- 173\relax
	\mciteBstWouldAddEndPuncttrue
	\mciteSetBstMidEndSepPunct{\mcitedefaultmidpunct}
	{\mcitedefaultendpunct}{\mcitedefaultseppunct}\relax
	\EndOfBibitem
	\bibitem[Walasik \latin{et~al.}(2017)Walasik, Silahli, and
	Litchinitser]{Walasik2017}
	Walasik,~W.; Silahli,~S.~Z.; Litchinitser,~N.~M. \emph{Sci. Rep.}
	\textbf{2017}, \emph{7}, 11709\relax
	\mciteBstWouldAddEndPuncttrue
	\mciteSetBstMidEndSepPunct{\mcitedefaultmidpunct}
	{\mcitedefaultendpunct}{\mcitedefaultseppunct}\relax
	\EndOfBibitem
	\bibitem[Feit and Fleck(1978)Feit, and Fleck]{Feit:78}
	Feit,~M.~D.; Fleck,~J.~A. \emph{Appl. Opt.} \textbf{1978}, \emph{17},
	3990--3998\relax
	\mciteBstWouldAddEndPuncttrue
	\mciteSetBstMidEndSepPunct{\mcitedefaultmidpunct}
	{\mcitedefaultendpunct}{\mcitedefaultseppunct}\relax
	\EndOfBibitem
	\bibitem[Lax \latin{et~al.}(1981)Lax, Batteh, and
	Agrawal]{doi:10.1063/1.328442}
	Lax,~M.; Batteh,~J.~H.; Agrawal,~G.~P. \emph{J. Appl. Phys.} \textbf{1981},
	\emph{52}, 109--125\relax
	\mciteBstWouldAddEndPuncttrue
	\mciteSetBstMidEndSepPunct{\mcitedefaultmidpunct}
	{\mcitedefaultendpunct}{\mcitedefaultseppunct}\relax
	\EndOfBibitem
\end{mcitethebibliography}

\providecommand{\latin}[1]{#1}
\providecommand*\mcitethebibliography{\thebibliography}
\csname @ifundefined\endcsname{endmcitethebibliography}
{\let\endmcitethebibliography\endthebibliography}{}

\end{document}